\documentclass[3p,twocolumn,authoryear,nonatbib]{elsarticle}
\usepackage[utf8]{inputenc}
\usepackage[T1]{fontenc}
\usepackage{csquotes}
\usepackage{amsmath,amssymb}
\usepackage{bm}
\usepackage{siunitx}
\usepackage{microtype}
\usepackage{ragged2e}
\usepackage[svgnames]{xcolor}
\usepackage[colorlinks=true,citecolor=Blue,allcolors=Blue,pdfencoding=auto, psdextra=true]{hyperref}
\usepackage{tikz}
\usepackage{caption,subcaption}
\usepackage{graphicx}
\usepackage{pgf}
\providecommand{\mathdefault}[1][]{}
\usepackage{subcaption}
\usepackage{booktabs}
\usepackage{adjustbox}
\usepackage{placeins}
\usepackage{stfloats}
\usepackage{fix-cm}
\usepackage{textcomp}
\usepackage{float}
\usepackage{comment}
\AtBeginDocument{%
  \def\@@author[#1]#2{%
    \g@addto@macro\elsprelimauthors{%
      \prelimauthorsep#2%
      \def\prelimauthorsep{\unskip,\space}}%
    \g@addto@macro\elsauthors{%
      \def\baselinestretch{1}%
      \authorsep{\large#2}\unskip\textsuperscript{%
        \@for\@@affmark:=#1\do{%
          \edef\affnum{\@ifundefined{X@\@@affmark}{1}{\elsRef{\@@affmark}}}%
          \unskip\sep\affnum\let\sep=,}%
        \ifx\@fnmark\@empty\else\unskip\sep\@fnmark\let\sep=,\fi
        \ifx\@corref\@empty\else\unskip\sep\@corref\let\sep=,\fi
        }%
      \def\authorsep{\unskip,\space}%
      \global\let\sep\@empty\global\let\@corref\@empty
      \global\let\@fnmark\@empty}%
    \@eadauthor={#2}%
    \g@addto@macro\useauthors{#2; }%
  }%
}

\usepackage[
	backend=biber,
	style=numeric-comp,
	sorting=none,
	backref=false, % true
	giveninits=true,  % Force les initiales pour les prÃ©noms
	maxcitenames=99,
	mincitenames=99,
	maxbibnames=99,
	minbibnames=99,
	block=space
	]{biblatex}
\DeclareSourcemap{
	\maps[datatype=bibtex]{
		\map{
			\step[fieldset=urldate, null]
		}
	}
}
\AtEveryBibitem{%
	\hypertarget{bib:\thefield{entrykey}}{}%
}

\usepackage[
acronym,
nomain
]{glossaries-extra}

\setabbreviationstyle{long-short}
\makeglossaries

\newacronym[first={MED (MÃ©thode des Ã‰lÃ©ments Discrets)}]{med}{MED}{MÃ©thode des Ã‰lÃ©ments Discrets}

\setglossarystyle{list}

\DefineBibliographyStrings{french}{
	phdthesis = {ThÃ¨se de doctorat},
}

\begin{document}
%%----------------------------------------------------------------------
\abstracttitle{{\large Abstract}} % 
\begin{frontmatter}

%% ------ Titre ------
\title{\textbf{{\Large Generation of representative powder particle packing in 2D/3D:\\ which tool for which application?}}}

% generation of packing 2d/3d of particle powder : who can we generate RVE in 2d/3d ? which tool can we use?

%% ------ Auteurs ------
\author[cemef]{{Antoine Tainturier}\corref{cor1}} %   \scriptsize
\ead{antoine.tainturier@minesparis.psl.eu}
\cortext[cor1]{Corresponding author}

\author[fra]{{\large Louis Lemarquis}}

\author[saf]{{\large Victor Szczepan}}

\author[cemef]{{\large Marc Bernacki}}

%% ------ Affiliations ------
\affiliation[cemef]{
  organization={CEMEF -- Centre de Mise en Forme des Mat\'{e}riaux, Mines Paris -- PSL},
  addressline={1 rue Claude Daunesse},
  city={06904 Sophia Antipolis},
  country={France}}

\affiliation[fra]{
  organization={Framatome Centre Technique},
  addressline={Rue Louis Alphonse Poitevin},
  city={71380 Saint-Marcel},
  country={France}}

\affiliation[saf]{
  organization={Safran Tech},
  addressline={Rue des Jeunes Bois CS80112},
  city={78771 Magny-les-Hameaux},
  country={France}}

\vspace{-1.5cm}
%% ------ Abstract ------
\begin{abstract}
	Although dense sphere packings serve as the initial state for simulations in powder metallurgy, additive manufacturing and granular physics, the choice of a packing generator is rarely guided by a systematic benchmark. A representative packing must be (i)--(ii)~physically admissible (non-overlapping particles in gravitational equilibrium); (iii)~faithful to the target particle size distribution (PSD); (iv)~representative in relative density $\varphi$; and (v)~computationally affordable. Four open-source tools have been benchmarked, meeting (i)--(ii) by construction: the sequential D\&R (dropping-and-rolling) and its densified variant D\&R-ME, and the discrete element method (DEM) codes LAMMPS (gravity) and dp3D (isostatic compression). Across four configurations (2D/3D lognormal, 3D bimodal, and a 3D domain-size study), they are compared against an industrial MIM-grade powder, with PSD fidelity measured by the bin-width-independent Hellinger distance and $\varphi$ against the feedstock solid loading ($\varphi_{\mathrm{exp}} = 0.62$, by Archimedes' method). In 3D, the DEM codes reach the densest packings but run more than three orders of magnitude slower: for $\sim\!20\,000$ particles, D\&R shows a $9\,\%$ $\varphi$ shortfall relative to dp3D while running $1800\times$ faster. These idealised model packings yield application-driven tool-selection guidelines.
\end{abstract}

%% ------ Mots-cles ------
\begin{keyword}
sphere packing \sep representative volume element \sep discrete element method \sep dropping and rolling \sep additive manufacturing
\end{keyword}

\end{frontmatter}

%Exemple d'utilisation du glossaire : cf. en haut (ligne 81) => La mÃ©thode \gls{med} est utilisÃ©e pour le frittage.
\section{Introduction}
	At the mesoscopic scale, generating a realistic dense packing of spherical particles is a prerequisite for the numerical simulation of a wide range of manufacturing and physical processes. In powder metallurgy, the initial packing topology --- and its relative density in particular --- governs densification kinetics, the evolution of the interparticle contact network, and the spatial distribution of residual porosity, both in free sintering \cite{bruchon20113DFiniteElement, pinomuñoz2013Direct3DSimulation, paredes-goyes2025LevelSetDiscrete, wang2006ComputerModelingSimulation,hötzer2019PhasefieldSimulationSolida,parhami1998NetworkModelInitial} and in hot isostatic pressing (HIP) \cite{arzt1983PracticalApplicationsHotisostatic, helle1985HotisostaticPressingDiagrams, bouvard1987ModellingHotIsostatic, zouaghi2013ÉtudeCompactionIsostatique,zouaghi2012ModellingCompactionPhase}.
	The same requirement arises in powder-bed additive manufacturing: in laser powder bed fusion (LPBF), the packing fraction of the spread layer directly governs energy-coupling efficiency, melt pool geometry, and thermal gradients \cite{yang2022ValidatedDimensionlessScaling, haapa2023ValidationPowderLayering, lee2018DynamicSimulationPowder, lampitella2021DiscreteElementMethod, fouda2020StudyPowderSpreading}; in binder jetting (BJ), packing fraction and particle arrangement condition binder saturation and green-part integrity \cite{ramalingam2025BinderJetAdditive, mostafaei2021BinderJet3D}.
	Dense packings also serve as the initial configuration for molecular dynamics simulations of hard-sphere systems, colloidal glasses, and granular physics \cite{plimpton1995FastParallelAlgorithms, anderson2020HOOMDbluePythonPackage, lubachevsky1991DisksVsSpheres, donev2004ImprovingDensityJammed}. In geomechanics, particle-scale assemblies model soils, rock masses, and granular flows \cite{cundall1979DiscreteNumericalModel, bagi2005AlgorithmGenerateRandom, labra2009HighdensitySpherePacking, potyondy2004BondedparticleModelRock, zhu2007DiscreteParticleSimulation, SullivanParticulate-discrete-element-modelling2011}. In electrochemical energy-storage devices such as lithium-ion batteries and solid oxide fuel cells, random packings often represent the porous architecture of composite electrodes \cite{lin2024ParticlePackingElectrode, zhang2011RandompackingModelSolid, usseglio-viretta2018ResolvingDiscrepancyTortuosity}. In all these cases, the packing constitutes the representative volume element (RVE) \cite{kanit2003DeterminationSizeRepresentative,bargmann2018Generation3DRepresentative}, i.e.\ the geometry from which subsequent simulations are initialised. Yet, despite the breadth of these applications, the literature still lacks a unified framework for deciding whether a generated packing is fit to serve as an initial state. Since this RVE is the very configuration from which the downstream simulation departs, we propose five criteria as a common basis for discussing powder-packing generation; all must be satisfied simultaneously:
	\begin{itemize}
	  \item[(i)] \textbf{Hard-sphere constraint:} no two particles may overlap, a requirement inherent to the hard-sphere (3D) or hard-disk (2D) model on which the initial state of most subsequent simulations relies.
	  \item[(ii)] \textbf{Gravitational equilibrium:} a particle is in gravitational equilibrium when the contact forces balance both the net force and the net moment acting on it. For a frictional and cohesionless particle, this is equivalent to requiring that the vertical projection of its centre of mass lies within the convex hull of its contact points.
	  \item[(iii)] \textbf{Particle size distribution (PSD):} the size distribution of the generated assembly must match the target, whether expressed as a continuous probability law or a discrete histogram defined over size bins. 
	  \item[(iv)] \textbf{Relative density:} the packing fraction must match an experimental reference value for the target powder. The usual Random Close Packing (RCP) limit is protocol-dependent and ill-suited to cross-method comparison; a refined metric is introduced in Section~\ref{sec:relative-density}.
	  \item[(v)] \textbf{Computational cost:} the generation time must remain acceptable relative to the cost of the downstream simulation.
	\end{itemize}

	Conditions~(i) and~(ii) are frequently neglected by purely geometric algorithms, yet they are critical when the packing serves directly as the initial state of a mechanical or sintering simulation. Even when all five conditions~(i)--(v) are met, the resulting assemblies remain idealised \emph{model packings}: they represent a dense, gravitationally stable packing of the powder, whereas the actual bed formed in a given process --- e.g.\ the spread layer in LPBF --- additionally depends on process-specific deposition mechanics not captured here, so extrapolation of a generated packing to a particular process must be made with care.\\[0.2 cm]
	Packing generation methods can be grouped into three broad families \cite{bargmann2018Generation3DRepresentative}:
	\begin{itemize}
	  \item \textbf{Dynamic methods} integrate the physical equations of motion governing each particle under gravity, contact forces, friction, compression or other externally applied actions (e.g.\ an LPBF recoater) until equilibrium is reached.
	  \item \textbf{Sequential geometric methods} place particles one by one under a local stability rule depending on already-placed neighbours (e.g.\ dropping-and-rolling).
	  \item \textbf{Dense constructive methods} assemble particles through purely geometric rules --- \textit{collective inflation} \cite{lubachevsky1991DisksVsSpheres, donev2004ImprovingDensityJammed}, \textit{advancing-front} schemes \cite{bagi2005AlgorithmGenerateRandom, labra2009HighdensitySpherePacking, valera2015ModifiedAlgorithmGenerating}, or \textit{random sequential addition} and its derivatives \cite{torquato2002RandomHeterogeneousMaterials} --- without integrating any equation of motion: they routinely reach very high relative densities but, lacking a force balance, fail condition~(ii) by construction, while inflation-driven variants transiently violate condition~(i).
	\end{itemize}
	To the authors' knowledge, no open-source dense-constructive implementation enforcing both~(i) and~(ii) is presently available, so this entire family is excluded from the present comparison. The remaining two families --- dynamic and sequential geometric methods --- are assessed against conditions~(i)--(v) in Section~\ref{sec:methods}, where a representative open-source implementation is selected for each. The resulting comparison reveals significant differences depending on the target configuration and leads to practical recommendations for tool selection.
	
\section{Packing generation methods}
\label{sec:methods}
	Many algorithms have been proposed for generating dense sphere packings, spanning commercial, academic and open-source implementations \cite{bargmann2018Generation3DRepresentative}; open-source or academic-version tools were deliberately selected here so that their source code, parameter files, and outputs can be freely examined and independently replicated by the community. Only the two families satisfying conditions~(i) and~(ii) by construction --- \textit{dynamic methods} and \textit{sequential geometric methods} --- are retained; for each, a representative open-source or academic-version method is selected and detailed below.\\[0.2 cm]
	\textbf{$\bullet$ Dynamic methods}\\[0.2 cm]
	Each particle $i$ of mass $m_i$, described by its centre position $\mathbf{x}_i$ and angular velocity $\boldsymbol{\omega}_i$, obeys Newton's equations of motion:
	\begin{equation}
		m_i \ddot{\mathbf{x}}_i = \mathbf{f}_i^{\mathrm{c}} + m_i \mathbf{g}, \qquad
		I_i \dot{\boldsymbol{\omega}}_i = \boldsymbol{\tau}_i^{\mathrm{c}},
		\label{eq:dem}
	\end{equation}
	where $\mathbf{f}_i^{\mathrm{c}}$ and $\boldsymbol{\tau}_i^{\mathrm{c}}$ are the net contact force and torque, $I_i$ the moment of inertia, and $\mathbf{g}$ the gravitational acceleration. The normal contact force generally follows a Hertz--Mindlin elasto-frictional law; adhesion is accounted for by the Johnson--Kendall--Roberts (JKR) model \cite{johnson1971SurfaceEnergyContact} or the Derjaguin--Muller--Toporov (DMT) model \cite{derjaguin1975EffectContactDeformations}, depending on the value of Tabor's parameter $\mu_T = \left(R\, W^{2} / E^{*2} z_0^{3}\right)^{1/3}$ \cite{martin2008InfluenceAdhesionFriction}, where $R$ is the effective particle radius, $W$ the work of adhesion, $E^{*}$ the effective elastic modulus and $z_0$ the equilibrium contact distance: the JKR model applies when $\mu_T \gg 1$ (compliant, large spheres with high adhesion energy), the DMT model otherwise. Two parameters drive the contact physics: the friction coefficient $\mu$ and the adhesion parameter $W$. Throughout this work both are taken from \cite{lee2018DynamicSimulationPowder}; the resulting packing can be modulated by tuning the couple $(W, \mu)$.\\[0.2 cm]
	Two packing modes are considered. In gravity mode (LAMMPS \cite{plimpton1995FastParallelAlgorithms}), particles settle under gravity until the total kinetic energy falls below a predefined threshold, directly yielding a gravitationally equilibrated assembly. In jamming mode (dp3D \cite{martin2003StudyParticleRearrangement, martin2014ContributionModélisationFrittage}), the domain boundary contracts isotropically under periodic boundary conditions until the volumetric strain rate satisfies $\dot{\varepsilon}_v < \dot{\varepsilon}_{v,\mathrm{stop}}$. In both modes, the radius $r_i$ of each inserted particle is drawn from the target number-weighted PSD, discretised into size bins upstream of the simulation, which enforces condition~(iii) by construction.\\[0.2 cm]
	\textbf{$\bullet$ Sequential geometric methods}\\[0.2 cm]
	Laguerre--Vorono\"{i} tessellations have led to the emergence of many algorithms in this family --- e.g.\ Neper \cite{quey2022NeperFEPXProject}, whose weighted power cells approximate the target PSD but share faces so the resulting grains overlap (condition~(i) violated) and which is primarily a polycrystalline-microstructure generator rather than a sphere-packing tool; it is therefore not retained. The representative method of this family is the dropping-and-rolling (D\&R) algorithm \cite{hitti2013OptimizedDroppingRolling, ilin2016AdvancingLayerAlgorithm}. Particles are introduced sequentially with radii $r_i$ drawn from the number-weighted PSD, discretised into size bins upstream of the simulation: each sphere is dropped from above and rolls over previously settled particles until it reaches the position of lowest gravitational potential energy compatible with mechanical stability, satisfying condition~(ii) by construction. Formally, the placement of particle $i$ minimises its vertical coordinate under the hard-sphere non-overlap constraint with every previously placed particle:
	\begin{equation}
		\mathbf{x}_i^* = \arg\min_{\mathbf{x}} \, x_z
		\quad \text{s.t.} \quad
		\|\mathbf{x} - \mathbf{x}_j\| \geq r_i + r_j, \; \forall j < i,
		\label{eq:dr}
	\end{equation}
	where $x_z$ is the vertical coordinate of the centre of particle $i$. In 3D, a particle is declared stable when it contacts at least three neighbours whose convex hull encloses the vertical projection of its centre. The packing terminates after a fixed number of consecutive failed placement attempts $N_{\text{max}}$, i.e.\ when no new sphere can be inserted without exiting the domain. Once an initial D\&R packing is generated, the moving-enlarging variant (D\&R-ME) post-processes it by displacing each particle towards the barycentre of its Delaunay neighbours (with decreasing step sizes under non-overlap constraints) and subsequently inflating each radius by the minimum clearance to any Delaunay neighbour; it does not strictly preserve the target PSD but reaches higher packing fractions than the classical D\&R. Both variants are implemented in the Cimlib library (CEMEF, Mines Paris -- PSL) \cite{hitti2013OptimizedDroppingRolling, ilin2016AdvancingLayerAlgorithm, digonnet2007CimlibFullyParallel,bernacki2024KineticEquationsLevelset}; D\&R is also distributed through an academic version of the DIGIMU\textsuperscript{\textregistered} software, commercialised by Transvalor \cite{digimu-5.0-2025}.\\[0.2 cm]
	Consequently, four implementations are retained for benchmarking: D\&R / D\&R-ME for the sequential geometric family \cite{hitti2013OptimizedDroppingRolling,ilin2016AdvancingLayerAlgorithm}, and LAMMPS \cite{plimpton1995FastParallelAlgorithms} (gravity) / dp3D \cite{martin2003StudyParticleRearrangement, martin2014ContributionModélisationFrittage} (jamming) for the dynamic family.
\section{Methodology, metrics and simulation parameters}
\label{sec:results}

	The four retained implementations are evaluated on four test configurations:
	\begin{enumerate}
		\item a 2D lognormal packing using an experimental MIM-grade powder PSD (see Table~\ref{tab:params-powder}), serving as a lightweight benchmark;
		\item a 3D binary bimodal study, probing method robustness under controlled polydispersity;
		\item a 3D lognormal packing in a reference cubic domain (MIM-grade powder PSD);
		\item a 3D lognormal domain-size parametric study, assessing convergence with respect to the domain size $X^3$ (MIM-grade powder PSD).
	\end{enumerate}
	Throughout this work, the lognormal distribution corresponds to an experimentally measured industrial MIM-grade IN718 nickel-based superalloy powder (Table~\ref{tab:params-powder}); a comparable IN718 MIM feedstock built on the same powder size distribution has been characterised in the literature \cite{royer2016DevelopmentCharacterizationMetal}. Its solid loading in the feedstock is independently known and used as a powder-specific reference. Table~\ref{tab:params-powder} summarises the corresponding lognormal PSD parameters (number distribution). This 3D granulometric distribution is used directly as input to all configurations, including the 2D case~(1); note that a planar cross-section of a 3D packing would instead follow the Wicksell transform of the 3D distribution~\cite{Saltykov1958}, so no direct comparison between configurations~(1) and~(3)--(4) is intended on this basis.\\[0.2 cm]
	Consistent with the family-level discussion of Section~\ref{sec:methods}, the four retained tools satisfy conditions~(i) and~(ii) by construction; the comparison therefore targets conditions~(iii)--(v): PSD fidelity in both number and surface/volume distributions, relative density, and computation time. The domain dimensions are fixed a priori while the achieved relative density, and hence the actual particle count, emerges from the prescribed PSD and from each method's intrinsic behaviour, the natural framing for sequential generation and one that ensures comparison on identical absolute domain sizes.	All simulations were performed sequentially on a single workstation (Intel Core Ultra~7 155H) under identical hardware and software conditions.\\[0.2 cm]
	\textbf{$\bullet$ Metrics}\\[0.2 cm]
	PSD fidelity is assessed on the radius histograms via the Hellinger distance. The generated assembly is binned into $K$ classes of width $\Delta r$; let $p_k = n_k/N$ be the empirical number-fraction in bin $k$ (with $n_k$ particles and $N = \sum_k n_k$ total), and $q_k = \int_{r_k}^{r_k+\Delta r}f_{\mathrm{target}}(r)\,\mathrm{d}r$ the corresponding target probability obtained analytically from the prescribed lognormal (no parametric fit to the generated histogram):
	\begin{equation}
		H(p,q) = \sqrt{\,1 - \sum_{k=1}^{K} \sqrt{p_k\, q_k}\,}\,.
		\label{eq:hellinger}
	\end{equation}
	$H\!\in[0,1]$, with $H = 0$ if and only if $p = q$ and $H = 1$ if and only if the supports of $p$ and $q$ are disjoint. Because $p_k$ and $q_k$ are probabilities (not densities), $H$ is bin-width independent and remains comparable across cases with different discretisations. The geometric-mean weighting $\sqrt{p_k q_k}$ also makes $H$ sensitive to discrepancies in low-probability bins (distribution tails) --- a property that proves critical when comparing number-weighted and surface/volume-weighted distributions, where tail particles dominate the latter, whereas an $L^2$ norm on the probability density would concentrate the penalty on the dense central bins. In the following, $H$ is computed on number-weighted frequencies (the direct simulation input) and on surface-weighted ($r^2$) frequencies for 2D cases or volume-weighted ($r^3$) frequencies for 3D cases; for conciseness, only the Hellinger distance $H$ is reported and discussed, the $L^2$ norm being omitted from the comparison.\\[0.2 cm]
	Relative density is defined as the volume (resp.\ area in 2D) fraction occupied by the $N$ particles:\label{sec:relative-density}
	\begin{equation}
	  \varphi = \frac{1}{|\Omega|}\sum_{i=1}^{N} v_i, \qquad
	  v_i = \begin{cases} \pi r_i^2 & \text{(2D)} \\ \dfrac{4}{3}\pi r_i^3 & \text{(3D)} \end{cases}
	  \label{eq:density}
	\end{equation}
	where the sum holds because condition~(i) prevents overlap. The domain measure $|\Omega| = \prod_\alpha L_\alpha$, with $\alpha \in \{x,y\}$ (2D) or $\{x,y,z\}$ (3D), is prescribed for periodic-boundary methods; for sequential methods, each $L_\alpha$ can be computed from the bounding box of the $N$ particle centres:
	\begin{equation}
	  L_\alpha = \max_{i=1,\dots,N}(\alpha_i) - \min_{i=1,\dots,N}(\alpha_i), \quad \alpha_{\min} = \min_{i}(\alpha_i).
	  \label{eq:bbox}
	\end{equation}
	To mitigate boundary effects and enable fair comparison between periodic-boundary and sequential methods, $\varphi$ is evaluated on the central $90\,\%$ sub-box $\mathcal{B}_{90}$, defined by insetting each face (3D) or edge (2D) of the bounding box by $5\,\%$ of $L_\alpha$:
	\begin{equation}
	  \mathcal{B}_{90} = \prod_\alpha \bigl[\alpha_{\min} + 0.05\,L_\alpha,\;\alpha_{\min} + 0.95\,L_\alpha\bigr].
	  \label{eq:subbox}
	\end{equation}

	For each face $f$ of $\mathcal{B}_{90}$, let $d_{i,f}$ be the signed distance from the centre of particle $i$ to face $f$, taken {positive} when the centre lies inside $\mathcal{B}_{90}$. Particle $i$ protrudes through face $f$ whenever $d_{i,f} < r_i$. All particles whose centre lies inside $\mathcal{B}_{90}$ are retained ($\mathcal{S}_B = \{i \mid \mathbf{x}_i \in \mathcal{B}_{90}\}$); the protruding piece through each face is subtracted, giving the clipped measure
	\begin{equation}
		v_i^{\mathrm{clip}} = v_i - \sum_{f:\,d_{i,f}<r_i} \Delta v_{i,f},
	\end{equation}
	where, in 2D (circular segment):
	\begin{equation}
	  \Delta v_{i,f}^{\,\mathrm{2D}} = r_i^2\arccos\!\left(\frac{d_{i,f}}{r_i}\right) - d_{i,f}\sqrt{r_i^2 - d_{i,f}^2}\,,
	  \label{eq:cap2d}
	\end{equation}
	and in 3D (spherical cap of height $h_{i,f} = r_i - d_{i,f}$):
	\begin{equation}
	  \Delta v_{i,f}^{\,\mathrm{3D}} = \frac{\pi\, h_{i,f}^2\!\left(3r_i - h_{i,f}\right)}{3}\,.
	  \label{eq:cap3d}
	\end{equation}
	The relative density is then:
	\begin{equation}
	  \varphi = \frac{\displaystyle\sum_{i \in \mathcal{S}_B} v_i^{\mathrm{clip}}}{\Omega_{90}}, \qquad \Omega_{90} = \prod_\alpha 0.9\,L_\alpha.
	  \label{eq:phiB}
	\end{equation}
	This estimator accounts for the exact particle measure within the sub-domain and provides an objective basis for comparison across generation methods. In the asymptotic regime of a large particle count, $\varphi$ as defined by Eq.~\ref{eq:phiB} converges to the standard definition used to characterise the canonical Random Close Packing limit ($\varphi_{\text{RCP}} \approx 0.64$) reported in the literature for periodic-boundary settings in 3D, and it consistently recovers the analytical limits of the canonical ordered lattices of equal spheres --- the square and hexagonal packings in 2D, and the BCC and HCP packings in 3D --- providing a complementary verification of the estimator. In the following results, every reported $\varphi$ is accompanied by the number of particles $N_{P,B}$ effectively retained in its computation (i.e.\ those lying in $\mathcal{B}_{90}$), which reflects how representative the sub-box sample is of the full packing.\\[0.2 cm]
	\textbf{$\bullet$ Simulation parameters}\\[0.2 cm]
	The reference domain sizes are matched to the experimental MIM-grade powder PSD (Table~\ref{tab:params-powder}): each box is chosen large enough relative to the experimental radius range ($r_{\min} = 0.1$ to $r_{\max} = 50\,\si{\micro\meter}$) to host $\sim\!2\times10^4$ particles at the densities achieved by the four methods, while remaining computationally affordable. They are set to $\SI{600}{\micro\meter}\!\times\!\SI{600}{\micro\meter}$ (2D) and $\SI{133}{\micro\meter}\!\times\!\SI{133}{\micro\meter}\!\times\!\SI{133}{\micro\meter}$ (3D). For the bimodal study, the per-case particle count $N$ is instead adjusted to guarantee at least $N_L = 100$ large particles ($N \gtrsim N_L / P_{N,L}$, with $P_{N,L}$ the number-probability of the large-particle class), and the domain size adapts to the mixture composition to keep the volume fraction $X_L$ comparable across methods (see Section~\ref{sec:results-discussion}). \\[0.2 cm]
	%	\FloatBarrier
	\begin{table*}[!b]
	\centering
	\caption{Lognormal PSD parameters in log-radius space (number distribution) of the experimental industrial MIM-grade IN718 powder, used in the 2D lognormal, 3D lognormal and domain-size parametric configurations.}
	\label{tab:params-powder}
	\small
		\begin{tabular}{lccccccc}
		\hline
		Case & $\mu_N$ & $\sigma_N$ & $\bar{r}$ (\si{\micro\meter}) & $s_r$ (\si{\micro\meter}) & $r_{\min}$ (\si{\micro\meter}) & $r_{\max}$ (\si{\micro\meter}) & $\varphi_{\mathrm{exp}}$ \\
		\hline
		IN718 MIM-grade (exp.) & 0.30 & 1.70 & 1.30 & 0.68 & 0.1 & 50 & 0.62 \\
		\hline
		\end{tabular}
	\end{table*}
%	\FloatBarrier
	In Table~\ref{tab:params-powder}, $\mu_N$ and $\sigma_N$ are the parameters of the lognormal law in log-radius space, $\bar{r}$ and $s_r$ the mean radius and number-weighted standard deviation, and $r_{\min}$, $r_{\max}$ the bounds imposed during sampling and discretisation. The discretisation of the lognormal law is performed intrinsically by each method, so the number- and surface/volume-weighted radius histograms in the following sections are reported per method. The reference value $\varphi_{\mathrm{exp}} = 0.62$ (Table~\ref{tab:params-powder}) is the powder solid loading of the industrial MIM feedstock --- $62\,\%$ powder for $38\,\%$ binder by volume, an injection-optimised composition validated by Archimedes' method. Since the binder merely fills the inter-particle voids, this solid loading provides a powder-specific estimate of the achievable packing fraction against which the generated packings can be compared.\\[0.2 cm]
	For DEM simulations (LAMMPS and dp3D), two cases are considered. The reference case, denoted $*\_{\mathrm{Lee2018}}$, uses the contact parameters of \cite{lee2018DynamicSimulationPowder}: friction coefficient $\mu = 0.577$, coefficient of restitution $\alpha_{\text{damp}} = 0.8$, and adhesion parameter $W = 4$. A second case, denoted $*\_{\mathrm{NANF}}$ (No Adhesion, No Friction), sets $(W,\mu) = (0, 0)$ and yields the upper-bound density $\varphi_{\max}$ for each method. For D\&R simulations, packing terminates after $N_{\max} = 200$ consecutive failed placement attempts \cite{hitti2013OptimizedDroppingRolling}, chosen large enough for the achieved density to plateau; the moving-enlarging variant (D\&R-ME) applies the same criterion and then post-processes the result with the moving-enlarging procedure described above. These parameters govern the equilibrium packing fraction: increasing adhesion $W$ or friction $\mu$ promotes looser configurations, while reducing damping yields denser packings, providing a practical lever to target a prescribed $\varphi$.
\section{Results \& Discussion}
\label{sec:results-discussion}
\textbf{$\bullet$ 2D lognormal}\\[0.2 cm]
	D\&R and D\&R-ME are run in native 2D mode; the DEM tools LAMMPS and dp3D are operated in pseudo-2D --- a 3D simulation collapsed to a single-particle-layer thickness ($z = 0$). While pseudo-2D is not the native mode for DEM codes, the physical force balance inherent to DEM naturally enforces gravitational equilibrium~(ii), making it a well-suited proxy for 2D packing generation. Table~\ref{tab:2d-micro-block} summarises conditions~(iii)--(v): PSD fidelity ($H$), relative density ($\varphi$) and computation time; Fig.~\ref{fig:2d-micro-block} exhibits representative packings for all four methods alongside number- and surface-weighted PSD histograms.
	\IfFileExists{plot/2d_plot/block.tex}{%
	  % % Block generated by nD_regroup.py, table + figure combined into ONE float.
% Required packages in the article preamble:
%   \usepackage{graphicx}
%   \usepackage{pgf}
%   \usepackage{tikz}
%   \usepackage{booktabs}
%   \usepackage{caption}  (provides \captionof)
% Layout: a single figure*[p] forces table + figure together on a dedicated
%         page-of-floats, in code order (table first, then figure).

\begin{figure*}[p]
  \centering
  % ---------- TABLE ----------
  \captionof{table}{Comparative performance of methods for 2D dense particle packing generation: accuracy metrics and computational efficiency.}
  \label{tab:2d-micro-block}
  \small
  \begin{tabular}{lccccccc}
    \toprule
    Method & X (µm) & $N_{\mathrm{part}}$ & $H_N$ & $H_S$ & $\varphi$ & $N_{P,B}$ & Comp.\ time \\
    \midrule
    D\&R & 600 & 20 654 & 0.030 & 0.14 & 0.811 & 16 546 & \SI{4.65}{\second} \\
    D\&R-ME & 600 & 20 694 & 0.036 & 0.14 & 0.823 & 16 578 & \SI{32.2}{\minute} \\
    LAMMPS\_Lee2018 & 600 & 25 557 & 0.040 & 0.12 & 0.918 & 21 091 & \SI{10.9}{\hour} \\
    LAMMPS\_NANF & 600 & 25 555 & 0.040 & 0.13 & 0.918 & 21 090 & \SI{10.9}{\hour} \\
    dp3D\_Lee2018 & 600 & 23 041 & 0.020 & 0.088 & 0.894 & 18 571 & \SI{18.0}{\hour} \\
    dp3D\_NANF & 600 & 23 118 & 0.020 & 0.088 & 0.901 & 18 653 & \SI{19.9}{\hour} \\
    \bottomrule
  \end{tabular}

  \vspace{1.2em}

  % ---------- FIGURE: 6 minipages (a)-(f) ----------
  \begin{minipage}[t]{0.495\textwidth}
    \centering
    \begin{tikzpicture}
      \node[inner sep=0pt] (img) {\resizebox{0.95\linewidth}{!}{\input{plot/2d_plot/DR_packing_inset.pgf}}};
      \node[anchor=north west, fill=white, fill opacity=0.65,
            text opacity=1, inner sep=-10pt, font=\large\bfseries]
        at (img.north west) {(a)};
    \end{tikzpicture}
  \end{minipage}\hfill
  \begin{minipage}[t]{0.495\textwidth}
    \centering
    \begin{tikzpicture}
      \node[inner sep=0pt] (img) {\resizebox{0.95\linewidth}{!}{\input{plot/2d_plot/DR-ME_packing_inset.pgf}}};
      \node[anchor=north west, fill=white, fill opacity=0.65,
            text opacity=1, inner sep=-10pt, font=\large\bfseries]
        at (img.north west) {(b)};
    \end{tikzpicture}
  \end{minipage}

  \begin{minipage}[t]{0.495\textwidth}
    \centering
    \begin{tikzpicture}
      \node[inner sep=0pt] (img) {\resizebox{0.95\linewidth}{!}{\input{plot/2d_plot/dp3D_packing_inset.pgf}}};
      \node[anchor=north west, fill=white, fill opacity=0.65,
            text opacity=1, inner sep=-10pt, font=\large\bfseries]
        at (img.north west) {(c)};
    \end{tikzpicture}
  \end{minipage}\hfill
  \begin{minipage}[t]{0.495\textwidth}
    \centering
    \begin{tikzpicture}
      \node[inner sep=0pt] (img) {\resizebox{0.95\linewidth}{!}{\input{plot/2d_plot/LAMMPS_packing_inset.pgf}}};
      \node[anchor=north west, fill=white, fill opacity=0.65,
            text opacity=1, inner sep=-10pt, font=\large\bfseries]
        at (img.north west) {(d)};
    \end{tikzpicture}
  \end{minipage}

  \vspace{0.5em}
  \begin{minipage}[t]{0.49\textwidth}
    \centering
    \begin{tikzpicture}
      \node[inner sep=0pt] (img) {\resizebox{\linewidth}{!}{\input{plot/2d_plot/article_number_linear.pgf}}};
      \node[anchor=north west, fill=white, fill opacity=0.65,
            text opacity=1, inner sep=-10pt, font=\large\bfseries]
        at (img.north west) {(e)};
    \end{tikzpicture}
  \end{minipage}\hfill
  \begin{minipage}[t]{0.49\textwidth}
    \centering
    \begin{tikzpicture}
      \node[inner sep=0pt] (img) {\resizebox{\linewidth}{!}{\input{plot/2d_plot/article_surface_linear.pgf}}};
      \node[anchor=north west, fill=white, fill opacity=0.65,
            text opacity=1, inner sep=-10pt, font=\large\bfseries]
        at (img.north west) {(f)};
    \end{tikzpicture}
  \end{minipage}

  \captionof{figure}{2D particle packings: spatial distributions (a)--(d) and particle size histograms, number-weighted (e), surface-weighted (f).}
  \label{fig:2d-micro-block}
\end{figure*}
	}{%
	  \textbf{Missing file:} \texttt{plot/2d_plot/block.tex}%
	}
	PSD fidelity differs significantly between methods. Despite the high polydispersity (coefficient of variation $s_r/\bar{r} \approx 0.52$), number-weighted histograms appear broadly consistent across all methods (Fig.~\ref{fig:2d-micro-block} (e)). The surface-weighted distribution reveals a weakness of LAMMPS (Fig.~\ref{fig:2d-micro-block} (f)): the largest generated radius reaches $r \approx 9\,\si{\micro\meter}$ against the prescribed $r_{\max} = 50\,\si{\micro\meter}$, irrespective of whether a continuous law or a discrete probability table is supplied as input. This systematic tail truncation originates in the pre-processing stage --- the discretisation of the continuous or discrete input PSD supplied to LAMMPS --- and not in the DEM contact resolution of Eq.~\ref{eq:dem} itself; it is therefore a limitation of the LAMMPS implementation, not of the DEM method. As a result, the LAMMPS relative density is only loosely comparable to the other three methods. D\&R, D\&R-ME and dp3D reproduce both weighted distributions faithfully, as confirmed by the $H$ values in Table~\ref{tab:2d-micro-block}.\\[0.2 cm]
	Regarding packing density, D\&R and D\&R-ME produce dense configurations efficiently; dp3D achieves a higher $\varphi$ at substantially greater computational cost --- in 2D, D\&R is roughly four orders of magnitude faster than dp3D (Table~\ref{tab:2d-micro-block}) for a $\sim\!10\,\%$ gain in $\varphi$ at $\sim\!2\times10^{4}$ particles. The DEM contact parameters of \cite{lee2018DynamicSimulationPowder} have a limited effect on the 2D packing fraction ($\Delta\varphi \approx 1\,\%$), though larger values of the pair $(W,\mu)$ could widen this gap.\\[0.2 cm]
\textbf{$\bullet$ 3D bimodal}\\[0.2 cm]
	A 3D binary bimodal study probes method robustness under polydispersity: the size ratio $\rho = R_L/R_S$ is varied over $\rho \in \{1, 2, 5, 10, 20, 40\}$ wherein $R_S = 1$ at a large-particle volume fraction $X_L = 0.75$, near the Furnas optimum \cite{furnas1931GradingAggregatesMathematical}; \textcite{meng2014PackingPropertiesBinary} locate the maximum packing density of binary hard-sphere mixtures near $X_L \approx 0.70$. The domain size adjusts automatically with the mixture composition so that all methods operate under comparable volume fractions; the per-case particle count follows from the representativity criterion below. Applying the representativity criterion of Section~\ref{sec:results} ($N_L \geq 100$) yields the per-case particle counts reported in Table~\ref{tab:bimodal-3D}, together with the achieved $\varphi$ and computational cost. Throughout the bimodal study, dp3D and LAMMPS are run in their NANF configuration $(W,\mu) = (0,0)$ to isolate the geometric effect of polydispersity from the contact-parameter dependence; D\&R-ME, whose Delaunay-based densification post-processing is designed for continuous distributions, is not applicable to a strictly binary mixture and is excluded from this study.\\[0.2 cm]
%	\FloatBarrier
	\begin{figure*}[!ht]
		\centering
		% ---------- TABLE ----------
		\captionof{table}{Performance of D\&R, LAMMPS and dp3D for the 3D binary bimodal packing at the Furnas composition $X_L = 0.75$, as a function of the size ratio $\rho$: relative density and computational cost.}
		\label{tab:bimodal-3D}
		\small
		\begin{tabular}{@{} *{7}{c} @{}}
			\toprule
			{Method} & ${\rho}$ & ${N_P}$ & ${P_{N,L}}$ & ${N_L}$ & ${\varphi}$ & {Comp.\ time} \\
			\midrule
			D\&R & 1 & $3.00 \times 10^{4}$ & $1.0 \times 10^{0}$ & 30 000 & 0.570 & \SI{5}{\second} \\
			D\&R & 2 & $3.00 \times 10^{4}$ & $2.7 \times 10^{-1}$ & 8 182 & 0.595 & \SI{6}{\second} \\
			D\&R & 5 & $3.00 \times 10^{4}$ & $2.3 \times 10^{-2}$ & 703 & 0.632 & \SI{15}{\second} \\
			D\&R & 10 & $3.30 \times 10^{4}$ & $3.0 \times 10^{-3}$ & 100 & 0.720 & \SI{37}{\second} \\
			D\&R & 20 & $2.70 \times 10^{5}$ & $3.8 \times 10^{-4}$ & 100 & 0.751 & \SI{59}{\minute} \\
			D\&R & 40 & $2.14 \times 10^{6}$ & $4.7 \times 10^{-5}$ & 100 & (0.644) & (\SI{23.8}{\hour}) \\
			\midrule
			LAMMPS & 1 & $3.00 \times 10^{4}$ & $1.0 \times 10^{0}$ & 30 000 & 0.657 & \SI{3.2}{\hour} \\
			LAMMPS & 2 & $3.00 \times 10^{4}$ & $2.7 \times 10^{-1}$ & 8 182 & 0.670 & \SI{4.2}{\hour} \\
			LAMMPS & 5 & $3.00 \times 10^{4}$ & $2.3 \times 10^{-2}$ & 703 & 0.755 & \SI{5.5}{\hour} \\
			LAMMPS & 10--20--40 & \multicolumn{1}{c}{--} & \multicolumn{1}{c}{$< 3.0 \times 10^{-3}$} & \multicolumn{3}{c}{Insufficient discretisation resolution} \\
			\midrule
			dp3D & 1 & $3.00 \times 10^{4}$ & $1.0 \times 10^{0}$ & 30 000 & 0.664 & \SI{1.1}{\hour} \\
			dp3D & 2 & $3.00 \times 10^{4}$ & $2.7 \times 10^{-1}$ & 8 182 & 0.692 & \SI{1.3}{\hour} \\
			dp3D & 5 & $3.00 \times 10^{4}$ & $2.3 \times 10^{-2}$ & 703 & 0.758 & \SI{1.8}{\hour} \\
			dp3D & 10 & $3.30 \times 10^{4}$ & $3.0 \times 10^{-3}$ & 100 & 0.788 & \SI{2.2}{\hour} \\
			dp3D & 20--40 & \multicolumn{1}{c}{--} & \multicolumn{1}{c}{$< 3.0 \times 10^{-3}$} & \multicolumn{3}{c}{Memory-limited} \\
			\bottomrule
		\end{tabular}

		\vspace{1.2em}

		% ---------- FIGURE: 2 minipages (a)-(b) ----------
		\begin{minipage}[t]{0.49\textwidth}
			\centering
			\begin{tikzpicture}
				\node[inner sep=0pt] (img) {\resizebox{\linewidth}{!}{\input{plot/bimodal_plot/phi_r.pgf}}};
				\node[anchor=north west, fill=white, fill opacity=0.65,
				text opacity=1, inner sep=-10pt, font=\large\bfseries]
				at (img.north west) {(a)};
			\end{tikzpicture}
		\end{minipage}\hfill
		\begin{minipage}[t]{0.49\textwidth}
			\centering
			\begin{tikzpicture}
				\node[inner sep=0pt] (img) {\resizebox{\linewidth}{!}{\input{plot/bimodal_plot/phi_XL.pgf}}};
				\node[anchor=north west, fill=white, fill opacity=0.65,
				text opacity=1, inner sep=-10pt, font=\large\bfseries]
				at (img.north west) {(b)};
			\end{tikzpicture}
		\end{minipage}

		\captionof{figure}{3D binary bimodal packings at the Furnas composition $X_L = 0.75$. (a) Relative density $\varphi$ versus size ratio $\rho$ for D\&R, LAMMPS and dp3D; (b) $\varphi$ versus large-particle volume fraction $X_L$ (Furnas plane), compared with the reference values of \cite{farr2009ClosePackingDensity, meng2014PackingPropertiesBinary, kyrylyuk2010PercolationJammingRandom}.}
		\label{fig:bimodal-3D-Meng}
	\end{figure*}
%	\FloatBarrier
	The DEM methods become impractical at high polydispersity: dp3D is memory-limited for $\rho \geq 20$, while LAMMPS fails to resolve the large-particle class once its number-probability drops below $\sim\!10^{-3}$ ($\rho \geq 10$) --- the same discretisation-resolution limit already observed in the 2D and 3D lognormal cases. D\&R remains usable up to $\rho = 20$ ($\varphi = 0.751$); at $\rho = 40$, despite the particle count being inflated to $\sim\!2\times10^{6}$ by the representativity constraint (a $\sim\!1$~day run), the achieved density does not follow the expected monotonic increase; it drops to $\varphi = 0.644$ instead of exceeding the $\rho = 20$ value. This anomalous point, parenthesised in Table~\ref{tab:bimodal-3D}, is most likely a consequence of the fixed stopping criterion $N_{\max} = 200$ used by D\&R: at such an extreme size disparity, the algorithm exits well before each large sphere has been properly surrounded by its small-sphere cortege, leaving extended void regions around the large particles and lowering the effective $\varphi$ below the $\rho = 20$ value. The point is therefore reported but excluded from the trend analysis. Computational times follow the same trend as in the lognormal cases: at $\rho = 5$, where all three methods succeed, D\&R is about $420\times$ and $1300\times$ faster than dp3D and LAMMPS respectively.\\[0.2 cm]
	Figure~\ref{fig:bimodal-3D-Meng} reports $\varphi$ versus the size ratio $\rho$ at the Furnas composition $X_L = 0.75$~(a) and its placement in the Furnas plane through the dependence on the large-particle volume fraction $X_L$~(b). The relative density increases monotonically with $\rho$ and saturates towards the Furnas upper bound \cite{furnas1931GradingAggregatesMathematical} at large size disparity. The three methods respond consistently to increasing polydispersity, with LAMMPS and dp3D close to the reference values of \cite{farr2009ClosePackingDensity, meng2014PackingPropertiesBinary, kyrylyuk2010PercolationJammingRandom} and D\&R systematically below them (Fig.~\ref{fig:bimodal-3D-Meng}~(b)), an offset consistent with the additional gravitational-equilibrium constraint~(ii) and the clipped-sub-box estimator enforced here.\\[0.2 cm]
\textbf{$\bullet$ 3D lognormal}\\[0.2 cm]
	The 3D study uses the same PSD parameters as the 2D case (Table~\ref{tab:params-powder}); 3D is the most appropriate comparison framework since dp3D and LAMMPS are inherently 3D codes. Fig.~\ref{fig:3d-micro-block} (a)--(d) shows representative 3D packings together with the method-specific PSD histograms in (e)--(f); Table~\ref{tab:3d-micro-block} summarises conditions~(iii)--(v): PSD fidelity ($H$), relative density ($\varphi$) and computation time.
%	\FloatBarrier
	\IfFileExists{plot/3d_plot/block.tex}{%
		% % Block generated by nD_regroup.py, table + figure combined into ONE float.
% Required packages in the article preamble:
%   \usepackage{graphicx}
%   \usepackage{pgf}
%   \usepackage{tikz}
%   \usepackage{booktabs}
%   \usepackage{caption}  (provides \captionof)
% Layout: a single figure*[p] forces table + figure together on a dedicated
%         page-of-floats, in code order (table first, then figure).

\begin{figure*}[p]
  \centering
  % ---------- TABLE ----------
  \captionof{table}{Comparative performance of methods for 3D dense particle packing generation: accuracy metrics and computational efficiency.}
  \label{tab:3d-micro-block}
  \small
  \begin{tabular}{lccccccc}
    \toprule
    Method & X (µm) & $N_{\mathrm{part}}$ & $H_N$ & $H_V$ & $\varphi$ & $N_{P,B}$ & Comp.\ time \\
    \midrule
    D\&R & 133 & 19 615 & 0.030 & 0.232 & 0.636 & 13 214 & \SI{26}{\second} \\
    D\&R-ME & 133 & 19 701 & 0.038 & 0.243 & 0.639 & 13 163 & \SI{10.0}{\minute} \\
    LAMMPS\_Lee2018 & 133 & 28 407 & 0.038 & 0.234 & 0.668 & 20 610 & \SI{10.8}{\hour} \\
    LAMMPS\_NANF & 133 & 26 726 & 0.042 & 0.219 & 0.698 & 19 115 & \SI{12.2}{\hour} \\
    dp3D\_Lee2018 & 133 & 26 000 & 0.069 & 0.085 & 0.584 & 18 899 & \SI{12.8}{\hour} \\
    dp3D\_NANF & 133 & 26 000 & 0.069 & 0.085 & 0.723 & 18 847 & \SI{1.07}{\day} \\
    \bottomrule
  \end{tabular}

  \vspace{1.2em}

  % ---------- FIGURE: 6 minipages (a)-(f) ----------
  \begin{minipage}[t]{0.495\textwidth}
    \centering
    \begin{tikzpicture}
      \node[inner sep=0pt] (img) {\resizebox{1.000\linewidth}{!}{\input{plot/3d_plot/DR_packing_3d.pgf}}};
      \node[anchor=north west, fill=white, fill opacity=0.65,
            text opacity=1, inner sep=-10pt, font=\large\bfseries]
        at (img.north west) {(a)};
    \end{tikzpicture}
  \end{minipage}\hfill
  \begin{minipage}[t]{0.495\textwidth}
    \centering
    \begin{tikzpicture}
      \node[inner sep=0pt] (img) {\resizebox{1.000\linewidth}{!}{\input{plot/3d_plot/DR-ME_packing_3d.pgf}}};
      \node[anchor=north west, fill=white, fill opacity=0.65,
            text opacity=1, inner sep=-10pt, font=\large\bfseries]
        at (img.north west) {(b)};
    \end{tikzpicture}
  \end{minipage}

  \vspace{0.4em}
  \begin{minipage}[t]{0.495\textwidth}
    \centering
    \begin{tikzpicture}
      \node[inner sep=0pt] (img) {\resizebox{1.000\linewidth}{!}{\input{plot/3d_plot/dp3D_packing_3d.pgf}}};
      \node[anchor=north west, fill=white, fill opacity=0.65,
            text opacity=1, inner sep=-10pt, font=\large\bfseries]
        at (img.north west) {(c)};
    \end{tikzpicture}
  \end{minipage}\hfill
  \begin{minipage}[t]{0.495\textwidth}
    \centering
    \begin{tikzpicture}
      \node[inner sep=0pt] (img) {\resizebox{1.000\linewidth}{!}{\input{plot/3d_plot/LAMMPS_packing_3d.pgf}}};
      \node[anchor=north west, fill=white, fill opacity=0.65,
            text opacity=1, inner sep=-10pt, font=\large\bfseries]
        at (img.north west) {(d)};
    \end{tikzpicture}
  \end{minipage}

  \vspace{0.5em}
  \begin{minipage}[t]{0.49\textwidth}
    \centering
    \begin{tikzpicture}
      \node[inner sep=0pt] (img) {\resizebox{\linewidth}{!}{\input{plot/3d_plot/article_number_linear_3D.pgf}}};
      \node[anchor=north west, fill=white, fill opacity=0.65,
            text opacity=1, inner sep=-10pt, font=\large\bfseries]
        at (img.north west) {(e)};
    \end{tikzpicture}
  \end{minipage}\hfill
  \begin{minipage}[t]{0.49\textwidth}
    \centering
    \begin{tikzpicture}
      \node[inner sep=0pt] (img) {\resizebox{\linewidth}{!}{\input{plot/3d_plot/article_volume_linear_3D.pgf}}};
      \node[anchor=north west, fill=white, fill opacity=0.65,
            text opacity=1, inner sep=-10pt, font=\large\bfseries]
        at (img.north west) {(f)};
    \end{tikzpicture}
  \end{minipage}

  \captionof{figure}{3D particle packings: spatial distributions (a)--(d) and particle size histograms, number-weighted (e), volume-weighted (f).}
  \label{fig:3d-micro-block}
\end{figure*}
	}{%
		\textbf{Missing file:} \texttt{plot/3d_plot/block.tex}%
	}
%	\FloatBarrier
	The trends echo the 2D case (Table~\ref{tab:3d-micro-block}): for the volume-weighted PSD, dp3D achieves the best $H$ value, followed by LAMMPS, then D\&R and D\&R-ME. This same distribution exposes the LAMMPS tail truncation already observed in 2D --- no particle with $r > 11\,\si{\micro\meter}$ is generated --- producing marked deviations in both the probability density and the cumulative PSD curves. The four packings appear visually dense (Fig.~\ref{fig:3d-micro-block} (a)--(d)), with $\varphi_\text{dp3D} > \varphi_\text{LAMMPS} > \varphi_\text{D\&R-ME} > \varphi_\text{D\&R}$; the higher densities reached by dp3D and LAMMPS come at a cost roughly $1800 \times$ that of D\&R for a $\Delta \varphi = 9\%$ shortfall compared to dp3D\_NANF. To support these observations, a parametric study over $X \in \{30, 80, 110, 133, 150, 210\}\,\si{\micro\meter}$ is presented in Fig.~\ref{fig:parametric-analysis-3d}, reporting the evolution of conditions~(iii)--(v) with domain size $X$.
%	\FloatBarrier
	\begin{figure*}[h!]
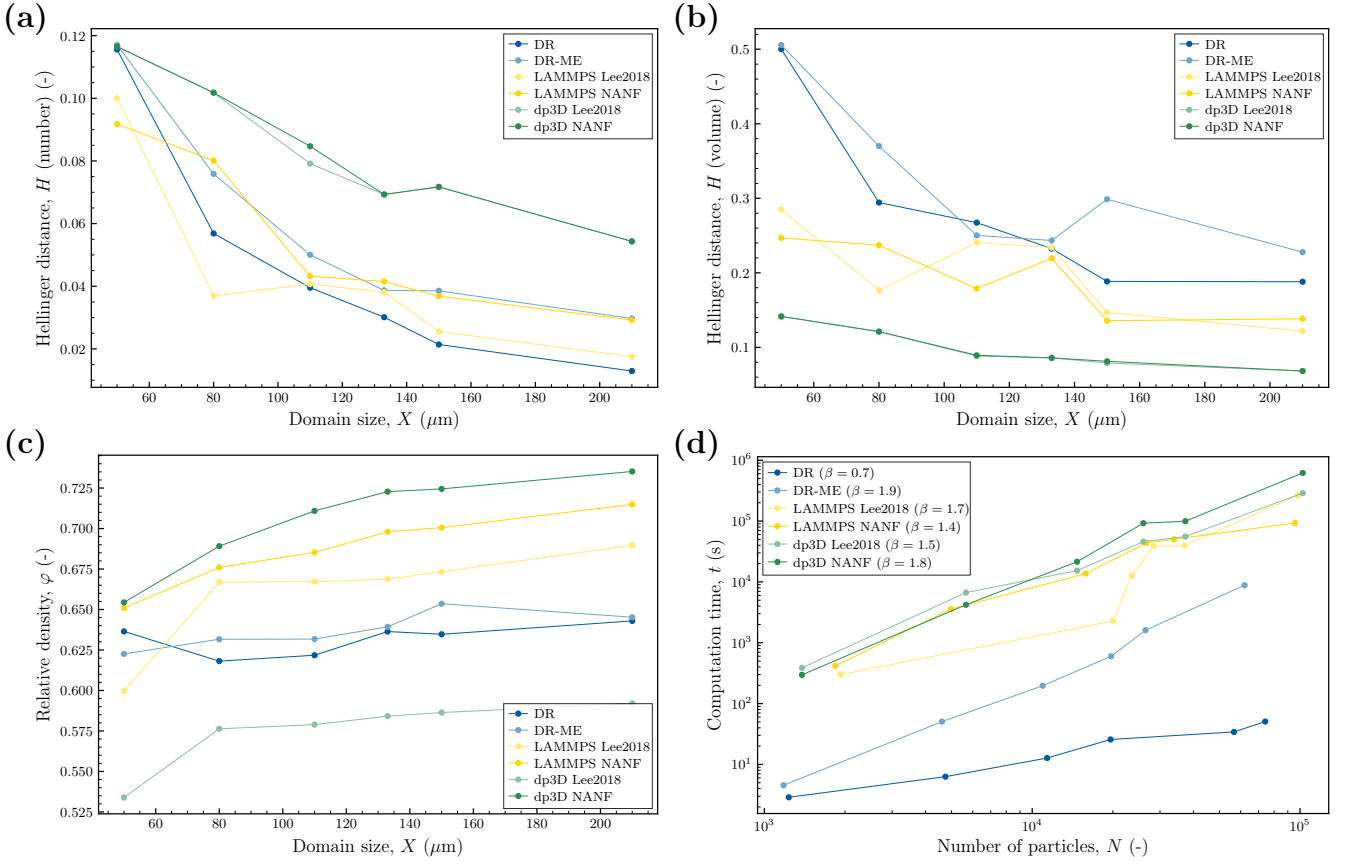

		\centering	
		\begin{minipage}[t]{0.49\textwidth}
			\centering
			\begin{tikzpicture}
				\node[inner sep=0pt] (img) {\resizebox{\linewidth}{!}{\input{plot/domain_plot/04a_hellinger_number_vs_x.pgf}}};
				\node[anchor=north west, fill=white, fill opacity=0.65,
				text opacity=1, inner sep=-10pt, font=\large\bfseries]
				at (img.north west) {(a)};
			\end{tikzpicture}
		\end{minipage}\hfill
		\begin{minipage}[t]{0.49\textwidth}
			\centering
			\begin{tikzpicture}
				\node[inner sep=0pt] (img) {\resizebox{\linewidth}{!}{\input{plot/domain_plot/05a_hellinger_volume_vs_x.pgf}}};
				\node[anchor=north west, fill=white, fill opacity=0.65,
				text opacity=1, inner sep=-10pt, font=\large\bfseries]
				at (img.north west) {(b)};
			\end{tikzpicture}
		\end{minipage}
		\vspace{0.5em}
		\begin{minipage}[t]{0.49\textwidth}
			\centering
			\begin{tikzpicture}
				\node[inner sep=0pt] (img) {\resizebox{\linewidth}{!}{\input{plot/domain_plot/03b_relative_density_b_vs_x.pgf}}};
				\node[anchor=north west, fill=white, fill opacity=0.65,
				text opacity=1, inner sep=-10pt, font=\large\bfseries]
				at (img.north west) {(c)};
			\end{tikzpicture}
		\end{minipage}\hfill
		\begin{minipage}[t]{0.49\textwidth}
			\centering
			\begin{tikzpicture}
				\node[inner sep=0pt] (img) {\resizebox{\linewidth}{!}{\input{plot/domain_plot/11_time_loglog_vs_particle_number.pgf}}};
				\node[anchor=north west, fill=white, fill opacity=0.65,
				text opacity=1, inner sep=-10pt, font=\large\bfseries]
				at (img.north west) {(d)};
			\end{tikzpicture}
		\end{minipage}
		\caption{Evolution of the Hellinger distance $H$, number-weighted (a) and volume-weighted (b), of the relative density $\varphi$ (c) and of the computational time (d) as functions of domain size and particle number in 3D granular packings.}
		\label{fig:parametric-analysis-3d}
	\end{figure*}
%	\FloatBarrier
	The evolution of $H$, $\varphi$ and $t$ with domain size $X$ follows the same trends as the fixed-domain comparison. As $X$ grows, the increasing particle count enables a finer discretisation of the target law, which results in a monotonic decrease of $H$ for both number- and volume-weighted distributions (Figs.~\ref{fig:parametric-analysis-3d} (a)--(b)). The relative density (Fig.~\ref{fig:parametric-analysis-3d} (c)) appears to converge towards an asymptotic plateau with the same method ordering as above: $\varphi_\text{dp3D} > \varphi_\text{LAMMPS} > \varphi_\text{D\&R-ME} > \varphi_\text{D\&R}$. The computation time (Fig.~\ref{fig:parametric-analysis-3d} (d)) obeys $t_\text{D\&R} < t_\text{D\&R-ME} < t_\text{LAMMPS} < t_\text{dp3D}$ across the entire domain range.\\[0.2 cm]
	The powder solid loading of the MIM feedstock, $\varphi_{\mathrm{exp}} = 0.62$ (Table~\ref{tab:params-powder}), provides a powder-specific target for the 3D lognormal case. D\&R and D\&R-ME reach $\varphi \approx 0.64$ (Table~\ref{tab:3d-micro-block}), within $\sim\!3\,\%$ of $\varphi_{\mathrm{exp}}$ without any parameter adjustment. For dp3D, $\varphi_{\mathrm{exp}}$ falls between the $*\_{\mathrm{Lee2018}}$ ($\varphi = 0.584$) and $*\_{\mathrm{NANF}}$ ($\varphi = 0.723$) bounds, so the compression increment --- or, equivalently, the adhesion--friction couple $(W,\mu)$ --- can be tuned to reproduce the experimental density. LAMMPS, by contrast, overshoots $\varphi_{\mathrm{exp}}$ in both configurations ($\varphi = 0.668$ and $0.698$ at $X = 133\,\si{\micro\meter}$); matching the experimental density would require contact parameters looser than those of \cite{lee2018DynamicSimulationPowder}. The direct comparison of the generated packings against a measured powder density --- rather than a protocol-dependent RCP value --- allows the relevance of each packing to be assessed with respect to the target powder.\\[0.2 cm]
	Method-specific control of $\varphi$ also differs significantly. D\&R and D\&R-ME do not offer a direct lever to tune the relative density once the PSD is fixed. DEM methods, in contrast, can modulate $\varphi$. dp3D leverages its underlying isostatic-compression physics to target a prescribed $\varphi_i \in [\varphi_0,\,\varphi_{\max}]$ directly through the chosen compression increment $i$, with $\varphi_{\max}$ recovered at $(W, \mu) = (0, 0)$. LAMMPS provides only indirect control via the adhesion--friction couple $(W, \mu)$, without any a priori prescription of the couple yielding a target $\varphi$.\\[0.2 cm]
	Although a maximum radius $r_{\max} = 50\,\si{\micro\meter}$ is imposed, the high polydispersity of the distribution means that the largest size classes --- e.g.\ $r = 25\,\si{\micro\meter}$ for $k = 40$ bins --- carry a small number probability $P_N (r = 25)$ but a non-negligible surface- or volume-weighted contribution ($\propto r^d$, $d = 2$ or $3$). Large particles therefore dominate the effective geometric and mechanical response, which directly raises the question of both statistical representativity and RVE validity. Following \cite{kanit2003DeterminationSizeRepresentative}, the RVE is the smallest volume for which effective properties become domain-size-independent with negligible statistical variance; in the present setting, this can be interpreted as the convergence plateau in $H$, and possibly in $\varphi$, observed in Fig.~\ref{fig:parametric-analysis-3d}. Since dp3D in its $*\_{\mathrm{NANF}}$ configuration reaches a higher $\varphi$ than D\&R, the latter can also serve as a low-cost initialisation for a subsequent dp3D isostatic compression up to a target $\varphi_i$, substantially shortening the path to a dense packing compared to running dp3D from scratch.\\[0.2 cm]
	The computational scaling of each method is characterised by an empirical power law
	\begin{equation}
		t(N) = a\, N^\beta,
	\end{equation}
	where $t$ denotes the computation time, $N$ the number of particles, $a$ a method-dependent prefactor and $\beta$ the effective scaling exponent obtained from log-log regression; lower $\beta$ indicates improved scalability with system size. D\&R exhibits the most favourable scaling, while D\&R-ME displays the strongest dependence on $N$. The remaining methods lie in $\beta \in [1.57,\,1.83]$, intermediate between linear and quadratic scaling.
\section{Conclusion}
	Four sphere-packing tools --- D\&R, D\&R-ME, LAMMPS, and dp3D --- were compared under unified conditions across four test configurations: 2D lognormal, 3D binary bimodal, 3D lognormal and a 3D domain-size parametric study. The hard-sphere and gravitational-equilibrium requirements are satisfied by construction in all retained methods; the proposed comparison therefore focused on PSD fidelity, relative density $\varphi$ and computation time, with a clipped-volume estimator of $\varphi$ on a central $90\,\%$ sub-box introduced to enable consistent cross-method comparison.\\[0.2 cm]
	dp3D and LAMMPS methods produce the densest packings while preserving the target PSD, but their runtimes exceed those of the sequential geometric methods by at least three orders of magnitude. dp3D additionally enables direct prescription of $\varphi_i \in [\varphi_0,\,\varphi_{\max}]$ through its compression-increment formulation, whereas LAMMPS offers only indirect tuning via the adhesion--friction couple $(W,\mu)$ and suffers from a systematic large-particle truncation that disqualifies it for highly polydisperse distributions. D\&R recovers the target PSD accurately at a fraction of the DEM cost: in 2D ($\sim\!2.0\!\times\!10^4$ particles) it is about $1.4\!\times\!10^{4}\times$ faster than dp3D for a $\sim\!10\,\%$ shortfall in $\varphi$, and in 3D ($\sim\!2.5\!\times\!10^4$ particles) the speed-up reaches $1800\times$ for a $9\,\%$ shortfall. The moving-enlarging variant D\&R-ME trades strict PSD respect for a marginally higher density --- below $1\,\%$ gain over D\&R in the same 3D configurations. Alternatively, dp3D can be used as a post-processing densification step on top of a D\&R packing, reaching a higher $\varphi$ at a fraction of the cost of running dp3D from scratch.\\[0.2 cm]
	The optimal choice is therefore application-driven: D\&R for fast prototyping and large RVEs where moderate density is sufficient, dp3D/LAMMPS when both a prescribed dense $\varphi$ and a faithful PSD are required, and D\&R-ME as a low-cost route to a marginally higher density above D\&R when slight PSD shifts are acceptable. A natural next step is the validation of these packings against neighbourhood-based descriptors --- e.g.\ mean coordination number or radial distribution function --- extracted from sufficiently resolved experimental tomographies of the same powders, which would extend the present benchmark beyond bulk density to the contact topology itself. Extension to non-spherical particle morphologies and the coupling of packing generators with downstream sintering or LPBF solvers constitute further natural directions for future work.
%\section*{Glossary}
%\printglossary[type=\acronymtype, title={}]
\newpage
\textbf{Declaration of generative AI and AI-assisted technologies in the manuscript preparation process}\\[0.2 cm]
	During the preparation of this work, the authors used Claude Sonnet 4.6 and GPT 5.2 to rephrase and improve the readability of selected passages. After using these tools, the authors reviewed and edited the content as needed and take full responsibility for the content of the published article.\\[0.2 cm]
\textbf{Acknowledgements}\\[0.2 cm]
	This work is part of a PhD project jointly funded by Framatome and Safran Tech within the DIGIMU consortium. The authors gratefully acknowledge their financial and technical support. Computational resources were provided by CEMEF, Mines Paris -- PSL.\\[0.2 cm]
\textbf{Author contributions}\\[0.2 cm]
	A. Tainturier: Methodology, Software, Investigation, Data curation, Visualization, Writing -- original draft, review \& editing.\\[0.2 cm]
	L. Lemarquis: Writing -- review \& editing.\\[0.2 cm]
	V. Szczepan: Data curation, Writing -- review \& editing.\\[0.2 cm]
	M. Bernacki: Conceptualization, Software (D\&R and D\&R-ME methods), Writing -- review \& editing, Supervision, Funding acquisition.\\[0.2 cm]
\textbf{Declaration of competing interests}\\[0.2 cm]
	The authors declare that they have no known competing financial interests or personal relationships that could have appeared to influence the work reported in this paper.

{\vspace{-10pt}
	\renewcommand*{\bibfont}{{{\tiny}}}
	\setlength{\bibhang}{-28pt}
	\defbibenvironment{bibliography}
	{\list{\printtext[labelnumberwidth]{\printfield{labelprefix}\printfield{labelnumber}}}%
		{\setlength{\labelwidth}{\labelnumberwidth}%
			\setlength{\leftmargin}{\labelwidth}%
			\addtolength{\leftmargin}{\labelsep}%
			\setlength{\itemsep}{2pt}%
			\setlength{\parsep}{0pt}}}
	{\endlist}
	{\item}
	\printbibliography[title={{\large Bibliography}}]
}

\end{document}